\documentclass[acmlarge]{acmart-ewaf}

\usepackage{multirow}
\usepackage[caption = false]{subfig}

\renewcommand{\baselinestretch}{1.15}

\AtBeginDocument{%
  }

\setcopyright{cc}
\copyrightyear{2025}
\acmYear{2025}
\acmConference[EWAF'25]{Fourth European Workshop on Algorithmic Fairness}{June 30--July 02, 2025}{Eindhoven, NL}

\begin{document}

\title{Investigating Popularity Bias Amplification in Recommender Systems Employed in the Entertainment Domain}

\author{Dominik Kowald}
\affiliation{%
  \institution{Know Center Research GmbH \& Graz University of Technology}
  \city{Graz}
  \country{Austria}}
\email{dkowald@know-center.at}


\begin{abstract}
Recommender systems have become an integral part of our daily online experience by analyzing past user behavior to suggest relevant content in entertainment domains such as music, movies, and books. Today, they are among the most widely used applications of AI and machine learning. Consequently, regulations and guidelines for trustworthy AI, such as the European AI Act, which addresses issues like bias and fairness, are highly relevant to the design, development, and evaluation of recommender systems. One particularly important type of bias in this context is popularity bias, which results in the unfair underrepresentation of less popular content in recommendation lists. 
This work summarizes our research on investigating the amplification of popularity bias in recommender systems within the entertainment sector. Analyzing datasets from three entertainment domains, music, movies, and anime, we demonstrate that an item's recommendation frequency is positively correlated with its popularity. As a result, user groups with little interest in popular content receive less accurate recommendations compared to those who prefer widely popular items. 
Furthermore, this work contributes to a better understanding of the connection between recommendation accuracy, calibration quality of algorithms, and popularity bias amplification. 
\end{abstract}



\maketitle

\section{Introduction and Motivation}
Recommender systems have become one of the most prevalent applications of machine learning and AI, shaping our daily online experiences. They play a crucial role in helping users navigate vast and complex information spaces by identifying relevant content~\cite{burke2011recommender,jannach2016recommender}. Since their early implementations~\cite{resnick1994grouplens}, these systems have relied on analyzing past user behavior to construct user models and provide recommendations, not only for items like movies, music, and books, but also for social connections in online networks~\cite{kowald2013social,lacic2015utilizing,eirinaki2018recommender}. 

A variety of techniques have been used to build these user models, including traditional methods such as collaborative filtering (CF)~\cite{ekstrand2011collaborative}, content-based filtering~\cite{lops2010content}, and hybrid approaches~\cite{burke2002hybrid}, as well as more recent techniques based on latent representations (embeddings) and deep learning~\cite{chen2023deep}. The entertainment domain is one of the key areas where recommender systems are widely deployed, assisting users in discovering movies, music, books, and other media. 
Given the increasing adoption of recommender systems in both research and industry~\cite{jannach2022impact}, and their inherently human-centric nature, it is essential to consider existing regulations and requirements for trustworthy AI~\cite{di2022recommender}. Various institutions, including the European Commission, have defined trustworthiness through multiple dimensions, leading to regulatory frameworks such as the \textit{EU Artificial Intelligence Act}\cite{webaiact}, which emphasizes issues like bias and fairness in AI. These concerns are particularly relevant in the context of recommender systems, as highlighted by recent studies on trustworthy recommendation models\cite{fan2023trustworthy}. 
While bias and fairness in AI and machine learning have received significant research attention in recent years~\cite{mehrabi2021survey,scher2023modelling}, the replication and amplification of biases remain open challenges, especially in interactive systems~\cite{friedman1996bias} and recommender systems in particular~\cite{chen2023bias}. One of the most prevalent biases in CF-based recommender systems is popularity bias, which results in the underrepresentation of less popular content in personalized recommendations~\cite{elahi2021investigating,abdollahpouri2021user}. 

The present work summarizes our research on investigating the amplification of popularity bias in recommender systems within the entertainment domain. Analyzing datasets from three sectors, music, movies, and animes, we demonstrate that popularity bias disproportionately impacts user groups with little interest in popular content, as they receive less accurate recommendations compared to those who favor popular items~\cite{kowald2020unfairness,kowald2021support,kowald2022popularity,ecir_bias_2023}. 
Furthermore, this work provides deeper insights into the relationship between recommendation accuracy, algorithmic calibration quality, and the amplification of popularity bias. Finally, we propose directions for future research.

\section{Background and Methods}
Research has demonstrated that recommendation algorithms, particularly those based on collaborative filtering (CF), exhibit a strong bias toward popular items, resulting in their overrepresentation in recommendation lists~\cite{ekstrand2018all,elahi2021investigating}. Conversely, this leads to the unfair underrepresentation of less popular, long-tail items~\cite{brynjolfsson2006niches,park2008long}. Various metrics have been proposed in the literature to assess and analyze popularity bias from both item and user perspectives~\cite{ahanger2022popularity,klimashevskaia2023survey}. In the following, we describe some of these metrics used in our research, and give an overview of datasets for recommender systems employed in the entertainment domain. 

\subsection{Metrics to Investigate Popularity Bias Amplification}
Our research focuses on three key methods for measuring popularity bias across user groups: (i) differences in recommendation accuracy, (ii) \textit{miscalibration}, and (iii) \textit{popularity lift}. The first approach involves a straightforward comparison of average recommendation accuracy between groups using the mean average error (\textit{MAE}). In contrast, \textit{miscalibration} and \textit{popularity lift} require more complex calculations.

In general, \textit{calibration} measures the alignment between a user profile $p$ and a corresponding recommendation list $q$ in terms of genre distribution~\cite{steck2018}. For example, if a user historically consumes 80\% rock and 20\% pop music, a \textit{calibrated} recommendation list should reflect a similar distribution. While not explicitly a popularity bias metric, calibration is frequently used to assess and interpret popularity bias in recommendations~\cite{abdollahpouri2019impact,abdollahpouri2020connection}. The concept of \textit{miscalibration} represents the deviation between $p$ and $q$~\cite{lin2020}, quantified using the \textit{Kullback-Leibler (KL)} divergence between the genre distributions in $p$, i.e., $p(c|u)$, and in $q$, i.e., $q(c|u)$:
\begin{align}
KL(p||q) = \sum_{c \in C} p(c|u) \log \frac{p(c|u)}{q(c|u)}
\end{align}
where $C$ is the set of all genres in the dataset. A value of $KL(p||q) = 0$ indicates perfect \textit{calibration}, while higher values (approaching 1) signify increasingly \textit{miscalibrated} recommendations. These values can be averaged for a given user group $g$. We term miscalibration as \textit{MC} in this paper. 

In contrast, \textit{popularity lift} quantifies the extent to which recommendation algorithms amplify the popularity bias inherent in user profiles~\cite{abdollahpouri2019unfairness,abdollahpouri2020connection}. Specifically, it measures the disproportionate recommendation of popular items to a given user group $g$. The metric is based on the group average popularity $GAP_p(g)$, which represents the mean popularity of items in the user profiles $p$ of group $g$. Similarly, $GAP_q(g)$ denotes the average popularity of recommended items for users in $g$. Popularity lift is then defined as:
\begin{align}
PL(g) = \frac{GAP_q(g) - GAP_p(g)}{GAP_p(g)}
\end{align}
A value of $PL(g) > 0$ indicates that recommendations for group $g$ are skewed toward more popular items, whereas $PL(g) < 0$ suggests an overrepresentation of less popular content. The ideal scenario is $PL(g) = 0$, where the popularity distribution remains unchanged. We term popularity lift as \textit{PL} in this paper. 

\subsection{Datasets and Recommender System Algorithms Employed in the Entertainment Domain}
In our research, we analyze three datasets from the entertainment section, namely Last.fm representing the music domain, MovieLens representing the movie domain, and MyAnimeList, representing the anime domain. These datasets are described in more detail in our previous work~\cite{kowald2022popularity,ecir_bias_2023}, are summarized in Table~\ref{tab:datasets}, and are freely available via Zenodo\footnote{\url{https://zenodo.org/records/7428435}}. As described in~\cite{kowald2022popularity,ecir_bias_2023}, we split the users in every dataset in three equally-sized groups (1,000 users) based on their inclination towards popularity. We term the groups \textit{LowPop}, \textit{MedPop}, and \textit{HighPop}.  

\begin{table}[h]
\centering
\caption{Statistics of the datasets, including the number of users ($|U|$), items ($|I|$), ratings ($|R|$), and distinct genres ($|C|$), as well as sparsity, average interactions per user/item, and the rating range ($R$-range)~\cite{kowald2022popularity,ecir_bias_2023}.}
\resizebox{0.80\textwidth}{!}{
\begin{tabular}{l c c c c | c c c | c}
\toprule
\textbf{Dataset} & \boldmath$|U|$ & \boldmath$|I|$ & \boldmath$|R|$ & \boldmath$|C|$ & \boldmath$|R|/|U|$ & \boldmath$|R|/|I|$ & \textbf{Sparsity} & \boldmath$R$-range \\ 
\midrule
Last.fm  & 3,000  & 131,188  & 1,417,791  & 20  & 473  & 11  & 0.996  & [1–1,000] \\
MovieLens   & 3,000  & 3,667    & 675,610    & 18  & 225  & 184 & 0.938  & [1–5] \\
MyAnimeList  & 3,000  & 9,450    & 649,814    & 44  & 216  & 69  & 0.977  & [1–10] \\
\bottomrule
\end{tabular}}
\label{tab:datasets}
\end{table}

We analyze two well-known personalized recommendation algorithms used in the entertainment section, namely user-based, k-nearest-neighbor CF (UserKNN)~\cite{herlocker1999algorithmic} and non-negative matrix factorization (NMF)~\cite{luo2014efficient}. For the sake of reproducibility~\cite{semmelrock2023reproducibility,semmelrock2024reproducibility}, the implementation details of these algorithms and the complete source code to reproduce all of our research results are available via GitHub\footnote{\url{https://github.com/domkowald/FairRecSys}}.

\section{Results and Findings} \label{s:res}
\begin{figure}[h]
\centering
   \subfloat[UserKNN]{
      \includegraphics[width=.39\textwidth]{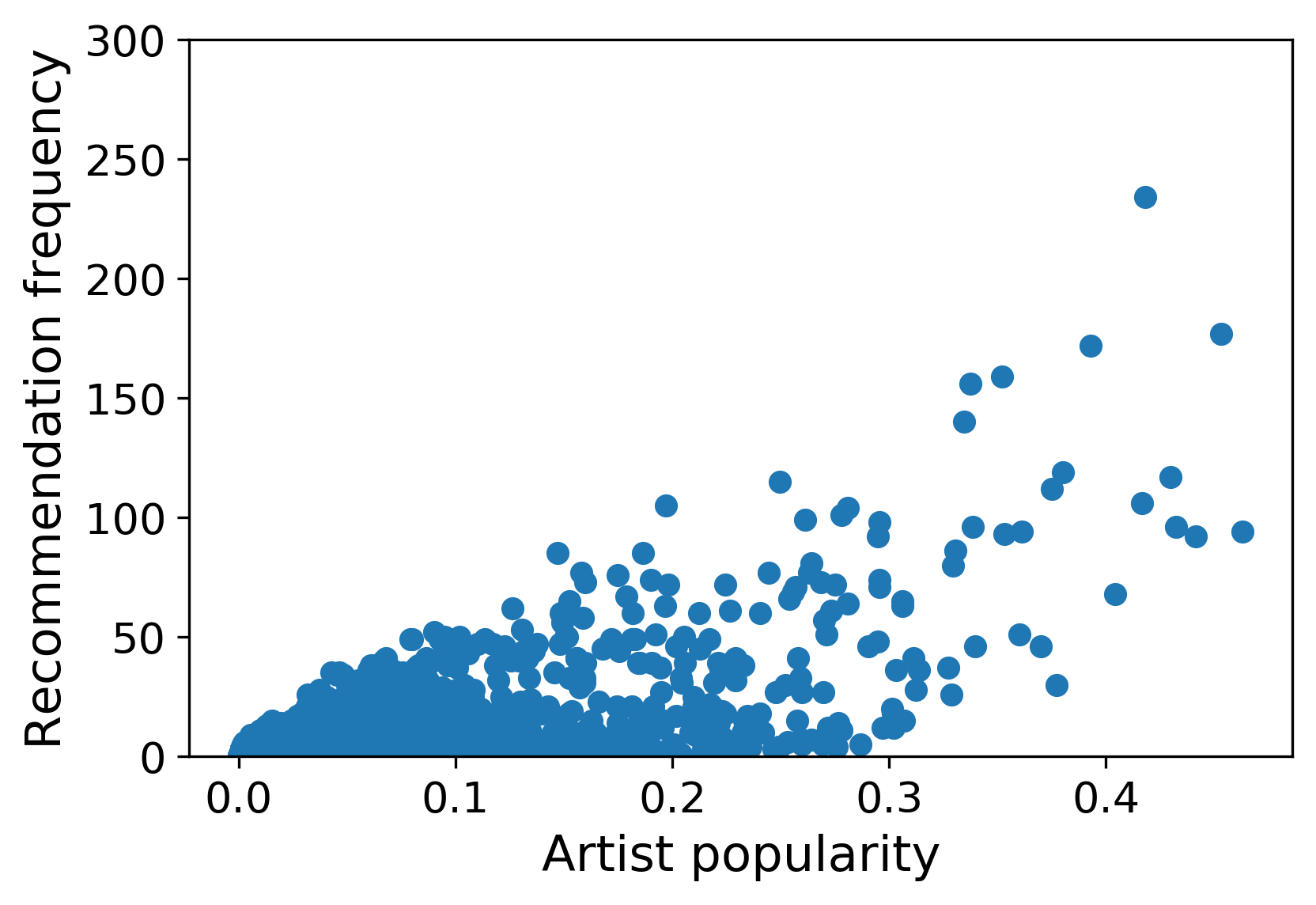}}
   \subfloat[NMF]{
      \includegraphics[width=.39\textwidth]{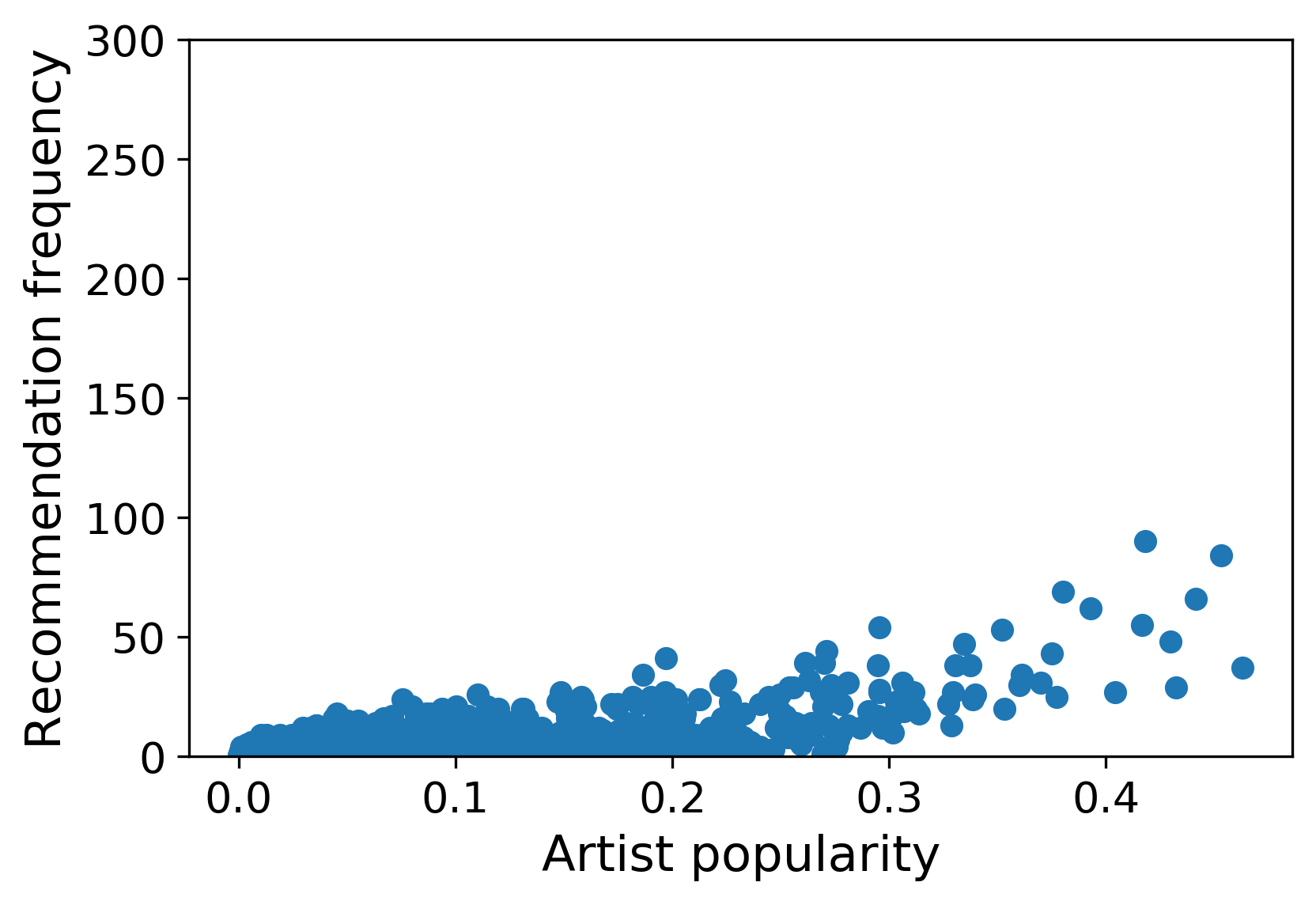}}
   \caption{Correlation of music artist popularity and recommendation frequency in the Last.fm dataset. Both algorithms investigated tend to favor popular music artists~\cite{kowald2020unfairness,kowald2021support}. Similar results can be obtained for the movie and anime domains~\cite{kowald2022popularity,ecir_bias_2023}\vspace{-8mm}}. 
   \label{fig:rec_popularity}
\end{figure}

In Figure~\ref{fig:rec_popularity}, we show the correlation between the popularity of music artists and their recommendation frequency in our Last.fm dataset. We see that both algorithms tend to favor popular music artists in their recommendation lists~\cite{kowald2020unfairness,kowald2021support}. For the sake of space, we omit the results for MovieLens and MyAnimeList, but similar results can be obtained for these domains as well~\cite{kowald2022popularity,ecir_bias_2023}. This means that the higher the popularity of an item, the higher is also the probability that this item is recommended. Next, we investigate if this item popularity bias also negatively influences user groups with little interest into popularity (i.e., our \textit{LowPop} group). 

Next, Table~\ref{tab:results} summarizes our results for the three entertainment datasets across the three user groups and the two algorithms. We see that the \textit{LowPop} user group always receives the statistically significant (according to a t-test with $p < 0.05$) worst accuracy results (\textit{MAE}). In addition, this user group also receives the most miscalibrated (\textit{MC}) and popularity biased (\textit{PL}) results, which helps to better understand the connection between accuracy and these two metrics~\cite{kowald2022popularity,ecir_bias_2023}.

Only in the Last.fm dataset, the other user groups reach worse \textit{PL} estimates, which we attribute to the special characteristics of this dataset, namely that the music domain is an entertainment domain with repeatedly consumed items~\cite{recsys_actr_2023,lex2020modeling,kowald2020utilizing}. Thus, in contrast to movies and animes, which are, in most cases, consumed only once or a few items, items in the music domain, such as artists or tracks, are typically consumed repeatedly. The \textit{PL} metric, however, only accounts for unique consumption patterns of items, and therefore this metric provides unintuitive results in the music domain. This suggest, that a weighted variant of \textit{PL} is needed, in which the popularity of an item for a given user is weighted with the number of items the user has consumed this item. We aim to implement and evaluate such an extended \textit{PL} metric in our future research, which we describe in the following section. 

\begin{table}[h]
    \centering
    \caption{\textit{MAE}, \textit{MC}, and \textit{PL} results for the \textit{LowPop}, \textit{MedPop}, and \textit{HighPop} user groups. The worst (i.e., highest) results are highlighted in \textbf{bold}. Statistical significance (t-test between \textit{LowPop} vs. \textit{MedPop} and \textit{HighPop}) is indicated by * for $p < 0.05$.}
    \label{tab:results}
    \resizebox{0.86\textwidth}{!}{
    \begin{tabular}{l l c c c | c c c | c c c}
    \toprule
    & \textbf{Data}    & \multicolumn{3}{c}{\textbf{Last.fm}}     & \multicolumn{3}{c}{\textbf{MovieLens}}     & \multicolumn{3}{c}{\textbf{MyAnimeList}}  \\ 
    \midrule
    \textbf{Algorithm}  & \textbf{Metric}   & \textit{MAE} & \textit{MC} & \textit{PL} & \textit{MAE} & \textit{MC} & \textit{PL} & \textit{MAE} & \textit{MC} & \textit{PL} \\ 
    \midrule
    \multirow{3}{*}{\textbf{UserKNN}} 
    & \textit{LowPop}  & \textbf{54.32}* & \textbf{0.51}* & 0.52 & \textbf{0.80}* & \textbf{0.75}* & \textbf{0.64}* & \textbf{1.37}* & \textbf{0.92}* & \textbf{0.74}*\\
    & \textit{MedPop}  & 46.76 & 0.50 & \textbf{0.82} & 0.75 & 0.69 & 0.37 & 1.34 & 0.72 & 0.22\\
    & \textit{HighPop}  & 49.75 & 0.45 & 0.80 & 0.72 & 0.62 & 0.20 & 1.31 & 0.63 & 0.08\\
    \midrule
    \multirow{3}{*}{\textbf{NMF}} 
    & \textit{LowPop}  & \textbf{42.47}* & \textbf{0.54}* & 0.10 & \textbf{0.75}* & \textbf{0.78}* & \textbf{0.57}* & \textbf{1.01}* & \textbf{0.91}* & \textbf{0.87}*\\
    & \textit{MedPop}  & 34.03 & 0.52 & 0.17 & 0.72 & 0.71 & 0.37 & 0.97 & 0.72 & 0.35\\
    & \textit{HighPop}  & 41.14 & 0.48 & \textbf{0.33} & 0.70 & 0.63 & 0.22 & 0.95 & 0.63 & 0.13\\
    \bottomrule
    \end{tabular}
    }
\end{table}

\section{Conclusion, Discussion, and Future Research Directions}
In this work, we have summarized our research on investigating the amplification of popularity bias in recommender systems within the entertainment sector. Analyzing datasets from music, movies, and anime, we have shown that recommendation frequency increases with item popularity, disadvantaging users with little interest in popular content. Additionally, we have explored the interplay between recommendation accuracy, algorithmic calibration, and the amplification of popularity bias. We want to further discuss this interplay and its relation to algorithmic fairness in the following paragraph. 

\paragraph{Discussion and Relation to Algorithmic Fairness} 
The metrics of accuracy differences, popularity lift, and miscalibration offer complementary perspectives on how recommender systems treat different user groups, and, together, they reveal important dimensions of algorithmic fairness. 
Accuracy differences expose disparities in the quality of recommendations received by different user groups. When users with low interest in popular content consistently receive lower accuracy scores, this suggests a systemic bias favoring mainstream preferences. This is not just a performance issue, but a fairness concern, as some users are consistently underserved.

Popularity lift further quantifies this issue by measuring the extent to which recommender systems amplify the popularity bias already present in user profiles. A high popularity lift for certain groups indicates that the system pushes them toward more popular items than they naturally prefer, potentially marginalizing niche tastes and reinforcing majority preferences. 
Miscalibration offers yet another lens, since it captures the mismatch between the user's historical preferences and the genre distribution in the recommendations. A high miscalibration value implies that the algorithm fails to reflect users’ true interests, which often coincides with lower accuracy and higher popularity lift, particularly for users with niche preferences.

Taken together, these three metrics provide a more holistic understanding of fairness. While accuracy differences reveal who is affected, popularity lift and miscalibration help explain how and why certain user groups are disadvantaged. A system that exhibits large disparities across all three dimensions can be considered unfair in its personalization strategy, as it systematically underrepresents the interests of less mainstream users. 
Thus, improving algorithmic fairness in recommender systems requires not only maximizing overall accuracy but also minimizing popularity lift and miscalibration, especially across diverse user groups. Balancing these metrics is essential for building systems that serve users equitably, beyond the majority preference.

\paragraph{Future Research Directions} 
Future research should investigate additional domains with respect to popularity bias amplification (e.g., news~\cite{lacic2022drives}), and also investigate the usefullness of the different metrics (especially popularity lift) in domains with repeat consumption patterns such as music, as already outlined in Section~\ref{s:res}.
Furthermore, we plan to work on robust methods for mitigating popularity bias, e.g., based on calibration techniques~\cite{klimashevskaia2022mitigating}, and evaluating these methods in online user studies (in addition to the offline experiments provided in this work). Finally, we would also like to investigate further aspects of trustworthiness in recommender systems, such as privacy, and study their relation to algorithmic fairness and popularity bias~\cite{mullner2024impact}.  

\begin{acks}
This work was supported by the FFG COMET program. A full version of this article was accepted as a habilitation at Graz University of Technology in June 2024, and is available via Arxiv~\citep{kowald2024transparency}.
\end{acks}

\newpage
\bibliographystyle{ACM-Reference-Format}
\bibliography{main_bibliography}


\begin{thebibliography}{45}


\ifx \showCODEN    \undefined \def \showCODEN     #1{\unskip}     \fi
\ifx \showDOI      \undefined \def \showDOI       #1{#1}\fi
\ifx \showISBNx    \undefined \def \showISBNx     #1{\unskip}     \fi
\ifx \showISBNxiii \undefined \def \showISBNxiii  #1{\unskip}     \fi
\ifx \showISSN     \undefined \def \showISSN      #1{\unskip}     \fi
\ifx \showLCCN     \undefined \def \showLCCN      #1{\unskip}     \fi
\ifx \shownote     \undefined \def \shownote      #1{#1}          \fi
\ifx \showarticletitle \undefined \def \showarticletitle #1{#1}   \fi
\ifx \showURL      \undefined \def \showURL       {\relax}        \fi
\providecommand\bibfield[2]{#2}
\providecommand\bibinfo[2]{#2}
\providecommand\natexlab[1]{#1}
\providecommand\showeprint[2][]{arXiv:#2}

\bibitem[Abdollahpouri et~al\mbox{.}(2019a)]%
        {abdollahpouri2019impact}
\bibfield{author}{\bibinfo{person}{Himan Abdollahpouri}, \bibinfo{person}{Masoud Mansoury}, \bibinfo{person}{Robin Burke}, {and} \bibinfo{person}{Bamshad Mobasher}.} \bibinfo{year}{2019}\natexlab{a}.
\newblock \showarticletitle{The impact of popularity bias on fairness and calibration in recommendation}.
\newblock \bibinfo{journal}{\emph{arXiv preprint arXiv:1910.05755}} (\bibinfo{year}{2019}).
\newblock


\bibitem[Abdollahpouri et~al\mbox{.}(2019b)]%
        {abdollahpouri2019unfairness}
\bibfield{author}{\bibinfo{person}{Himan Abdollahpouri}, \bibinfo{person}{Masoud Mansoury}, \bibinfo{person}{Robin Burke}, {and} \bibinfo{person}{Bamshad Mobasher}.} \bibinfo{year}{2019}\natexlab{b}.
\newblock \showarticletitle{The Unfairness of popularity bias in recommendation}. In \bibinfo{booktitle}{\emph{RMSE Workshop co-located with the 13th ACM Conference on Recommender Systems (RecSys 2019)}}.
\newblock


\bibitem[Abdollahpouri et~al\mbox{.}(2020)]%
        {abdollahpouri2020connection}
\bibfield{author}{\bibinfo{person}{Himan Abdollahpouri}, \bibinfo{person}{Masoud Mansoury}, \bibinfo{person}{Robin Burke}, {and} \bibinfo{person}{Bamshad Mobasher}.} \bibinfo{year}{2020}\natexlab{}.
\newblock \showarticletitle{The connection between popularity bias, calibration, and fairness in recommendation}. In \bibinfo{booktitle}{\emph{Fourteenth ACM Conference on Recommender Systems}}. \bibinfo{pages}{726--731}.
\newblock


\bibitem[Abdollahpouri et~al\mbox{.}(2021)]%
        {abdollahpouri2021user}
\bibfield{author}{\bibinfo{person}{Himan Abdollahpouri}, \bibinfo{person}{Masoud Mansoury}, \bibinfo{person}{Robin Burke}, \bibinfo{person}{Bamshad Mobasher}, {and} \bibinfo{person}{Edward Malthouse}.} \bibinfo{year}{2021}\natexlab{}.
\newblock \showarticletitle{User-centered evaluation of popularity bias in recommender systems}. In \bibinfo{booktitle}{\emph{Proceedings of the 29th ACM Conference on User Modeling, Adaptation and Personalization}}.
\newblock


\bibitem[Ahanger et~al\mbox{.}(2022)]%
        {ahanger2022popularity}
\bibfield{author}{\bibinfo{person}{Abdul~Basit Ahanger}, \bibinfo{person}{Syed~Wajid Aalam}, \bibinfo{person}{Muzafar~Rasool Bhat}, {and} \bibinfo{person}{Assif Assad}.} \bibinfo{year}{2022}\natexlab{}.
\newblock \showarticletitle{Popularity bias in recommender systems - a review}. In \bibinfo{booktitle}{\emph{International Conference on Emerging Technologies in Computer Engineering}}. Springer, \bibinfo{pages}{431--444}.
\newblock


\bibitem[Brynjolfsson et~al\mbox{.}(2006)]%
        {brynjolfsson2006niches}
\bibfield{author}{\bibinfo{person}{Erik Brynjolfsson}, \bibinfo{person}{Yu~Jeffrey Hu}, {and} \bibinfo{person}{Michael~D Smith}.} \bibinfo{year}{2006}\natexlab{}.
\newblock \showarticletitle{From niches to riches: Anatomy of the long tail}.
\newblock \bibinfo{journal}{\emph{Sloan Management Review}} \bibinfo{volume}{47}, \bibinfo{number}{4} (\bibinfo{year}{2006}), \bibinfo{pages}{67--71}.
\newblock


\bibitem[Burke(2002)]%
        {burke2002hybrid}
\bibfield{author}{\bibinfo{person}{Robin Burke}.} \bibinfo{year}{2002}\natexlab{}.
\newblock \showarticletitle{Hybrid recommender systems: Survey and experiments}.
\newblock \bibinfo{journal}{\emph{User Modeling and User-adapted Interaction}}  \bibinfo{volume}{12} (\bibinfo{year}{2002}).
\newblock


\bibitem[Burke et~al\mbox{.}(2011)]%
        {burke2011recommender}
\bibfield{author}{\bibinfo{person}{Robin Burke}, \bibinfo{person}{Alexander Felfernig}, {and} \bibinfo{person}{Mehmet~H G{\"o}ker}.} \bibinfo{year}{2011}\natexlab{}.
\newblock \showarticletitle{Recommender systems: An overview}.
\newblock \bibinfo{journal}{\emph{AI Magazine}} \bibinfo{volume}{32}, \bibinfo{number}{3} (\bibinfo{year}{2011}), \bibinfo{pages}{13--18}.
\newblock


\bibitem[Chen et~al\mbox{.}(2023a)]%
        {chen2023bias}
\bibfield{author}{\bibinfo{person}{Jiawei Chen}, \bibinfo{person}{Hande Dong}, \bibinfo{person}{Xiang Wang}, \bibinfo{person}{Fuli Feng}, \bibinfo{person}{Meng Wang}, {and} \bibinfo{person}{Xiangnan He}.} \bibinfo{year}{2023}\natexlab{a}.
\newblock \showarticletitle{Bias and debias in recommender system: A survey and future directions}.
\newblock \bibinfo{journal}{\emph{ACM Transactions on Information Systems}} \bibinfo{volume}{41}, \bibinfo{number}{3} (\bibinfo{year}{2023}), \bibinfo{pages}{1--39}.
\newblock


\bibitem[Chen et~al\mbox{.}(2023b)]%
        {chen2023deep}
\bibfield{author}{\bibinfo{person}{Xiaocong Chen}, \bibinfo{person}{Lina Yao}, \bibinfo{person}{Julian McAuley}, \bibinfo{person}{Guanglin Zhou}, {and} \bibinfo{person}{Xianzhi Wang}.} \bibinfo{year}{2023}\natexlab{b}.
\newblock \showarticletitle{Deep reinforcement learning in recommender systems: A survey and new perspectives}.
\newblock \bibinfo{journal}{\emph{Knowledge-Based Systems}}  \bibinfo{volume}{264} (\bibinfo{year}{2023}), \bibinfo{pages}{110335}.
\newblock


\bibitem[Commission(2021)]%
        {webaiact}
\bibfield{author}{\bibinfo{person}{European Commission}.} \bibinfo{year}{2021}\natexlab{}.
\newblock \bibinfo{title}{Proposal for a regulation laying down harmonised rules on Artificial Intelligence (EU AI Act), URL: https://eur-lex.europa.eu/legal-content/EN/TXT/?uri=CELEX:52021PC0206}.
\newblock
\newblock
\newblock
\shownote{Accessed on November 22nd, 2023}.


\bibitem[Di~Noia et~al\mbox{.}(2022)]%
        {di2022recommender}
\bibfield{author}{\bibinfo{person}{Tommaso Di~Noia}, \bibinfo{person}{Nava Tintarev}, \bibinfo{person}{Panagiota Fatourou}, {and} \bibinfo{person}{Markus Schedl}.} \bibinfo{year}{2022}\natexlab{}.
\newblock \showarticletitle{Recommender systems under European AI regulations}.
\newblock \bibinfo{journal}{\emph{Commun. ACM}} \bibinfo{volume}{65}, \bibinfo{number}{4} (\bibinfo{year}{2022}), \bibinfo{pages}{69--73}.
\newblock


\bibitem[Eirinaki et~al\mbox{.}(2018)]%
        {eirinaki2018recommender}
\bibfield{author}{\bibinfo{person}{Magdalini Eirinaki}, \bibinfo{person}{Jerry Gao}, \bibinfo{person}{Iraklis Varlamis}, {and} \bibinfo{person}{Konstantinos Tserpes}.} \bibinfo{year}{2018}\natexlab{}.
\newblock \showarticletitle{Recommender systems for large-scale social networks: A review of challenges and solutions}.
\newblock \bibinfo{journal}{\emph{Future Generation Computer Systems}}  \bibinfo{volume}{78} (\bibinfo{year}{2018}), \bibinfo{pages}{413--418}.
\newblock


\bibitem[Ekstrand et~al\mbox{.}(2011)]%
        {ekstrand2011collaborative}
\bibfield{author}{\bibinfo{person}{Michael~D Ekstrand}, \bibinfo{person}{John~T Riedl}, \bibinfo{person}{Joseph~A Konstan}, {et~al\mbox{.}}} \bibinfo{year}{2011}\natexlab{}.
\newblock \showarticletitle{Collaborative filtering recommender systems}.
\newblock \bibinfo{journal}{\emph{Foundations and Trends in Human--Computer Interaction}} \bibinfo{volume}{4}, \bibinfo{number}{2} (\bibinfo{year}{2011}), \bibinfo{pages}{81--173}.
\newblock


\bibitem[Ekstrand et~al\mbox{.}(2018)]%
        {ekstrand2018all}
\bibfield{author}{\bibinfo{person}{Michael~D Ekstrand}, \bibinfo{person}{Mucun Tian}, \bibinfo{person}{Ion~Madrazo Azpiazu}, \bibinfo{person}{Jennifer~D Ekstrand}, \bibinfo{person}{Oghenemaro Anuyah}, \bibinfo{person}{David McNeill}, {and} \bibinfo{person}{Maria~Soledad Pera}.} \bibinfo{year}{2018}\natexlab{}.
\newblock \showarticletitle{All the cool kids, how do they fit in?: Popularity and demographic biases in recommender evaluation and effectiveness}. In \bibinfo{booktitle}{\emph{Conference on Fairness, Accountability and Transparency}}. PMLR, \bibinfo{pages}{172--186}.
\newblock


\bibitem[Elahi et~al\mbox{.}(2021)]%
        {elahi2021investigating}
\bibfield{author}{\bibinfo{person}{Mehdi Elahi}, \bibinfo{person}{Danial~Khosh Kholgh}, \bibinfo{person}{Mohammad~Sina Kiarostami}, \bibinfo{person}{Sorush Saghari}, \bibinfo{person}{Shiva~Parsa Rad}, {and} \bibinfo{person}{Marko Tkal{\v{c}}i{\v{c}}}.} \bibinfo{year}{2021}\natexlab{}.
\newblock \showarticletitle{Investigating the impact of recommender systems on user-based and item-based popularity bias}.
\newblock \bibinfo{journal}{\emph{Information Processing \& Management}} \bibinfo{volume}{58}, \bibinfo{number}{5} (\bibinfo{year}{2021}).
\newblock


\bibitem[Fan et~al\mbox{.}(2023)]%
        {fan2023trustworthy}
\bibfield{author}{\bibinfo{person}{Wenqi Fan}, \bibinfo{person}{Xiangyu Zhao}, \bibinfo{person}{Lin Wang}, \bibinfo{person}{Xiao Chen}, \bibinfo{person}{Jingtong Gao}, \bibinfo{person}{Qidong Liu}, {and} \bibinfo{person}{Shijie Wang}.} \bibinfo{year}{2023}\natexlab{}.
\newblock \showarticletitle{Trustworthy recommender systems: Foundations and frontiers}. In \bibinfo{booktitle}{\emph{Proceedings of the 29th ACM SIGKDD Conference on Knowledge Discovery and Data Mining}}. \bibinfo{pages}{5796--5797}.
\newblock


\bibitem[Friedman and Nissenbaum(1996)]%
        {friedman1996bias}
\bibfield{author}{\bibinfo{person}{Batya Friedman} {and} \bibinfo{person}{Helen Nissenbaum}.} \bibinfo{year}{1996}\natexlab{}.
\newblock \showarticletitle{Bias in computer systems}.
\newblock \bibinfo{journal}{\emph{ACM Transactions on Information Systems (TOIS)}} \bibinfo{volume}{14}, \bibinfo{number}{3} (\bibinfo{year}{1996}), \bibinfo{pages}{330--347}.
\newblock


\bibitem[Herlocker et~al\mbox{.}(1999)]%
        {herlocker1999algorithmic}
\bibfield{author}{\bibinfo{person}{Jonathan~L Herlocker}, \bibinfo{person}{Joseph~A Konstan}, \bibinfo{person}{Al Borchers}, {and} \bibinfo{person}{John Riedl}.} \bibinfo{year}{1999}\natexlab{}.
\newblock \showarticletitle{An algorithmic framework for performing collaborative filtering}. In \bibinfo{booktitle}{\emph{Proceedings of the 22nd Annual International ACM SIGIR Conference on Research and Development in Information Retrieval}}.
\newblock


\bibitem[Jannach et~al\mbox{.}(2016)]%
        {jannach2016recommender}
\bibfield{author}{\bibinfo{person}{Dietmar Jannach}, \bibinfo{person}{Paul Resnick}, \bibinfo{person}{Alexander Tuzhilin}, {and} \bibinfo{person}{Markus Zanker}.} \bibinfo{year}{2016}\natexlab{}.
\newblock \showarticletitle{Recommender systems — beyond matrix completion}.
\newblock \bibinfo{journal}{\emph{Commun. ACM}} \bibinfo{volume}{59}, \bibinfo{number}{11} (\bibinfo{year}{2016}), \bibinfo{pages}{94--102}.
\newblock


\bibitem[Jannach and Zanker(2022)]%
        {jannach2022impact}
\bibfield{author}{\bibinfo{person}{Dietmar Jannach} {and} \bibinfo{person}{Markus Zanker}.} \bibinfo{year}{2022}\natexlab{}.
\newblock \showarticletitle{Impact and value of recommender systems}.
\newblock \bibinfo{journal}{\emph{Recommender Systems Handbook}} (\bibinfo{year}{2022}).
\newblock


\bibitem[Klimashevskaia et~al\mbox{.}(2022)]%
        {klimashevskaia2022mitigating}
\bibfield{author}{\bibinfo{person}{Anastasiia Klimashevskaia}, \bibinfo{person}{Mehdi Elahi}, \bibinfo{person}{Dietmar Jannach}, \bibinfo{person}{Christoph Trattner}, {and} \bibinfo{person}{Lars Skj{\ae}rven}.} \bibinfo{year}{2022}\natexlab{}.
\newblock \showarticletitle{Mitigating popularity bias in recommendation: Potential and limits of calibration approaches}. In \bibinfo{booktitle}{\emph{International Workshop on Algorithmic Bias in Search and Recommendation}}. Springer, \bibinfo{pages}{82--90}.
\newblock


\bibitem[Klimashevskaia et~al\mbox{.}(2023)]%
        {klimashevskaia2023survey}
\bibfield{author}{\bibinfo{person}{Anastasiia Klimashevskaia}, \bibinfo{person}{Dietmar Jannach}, \bibinfo{person}{Mehdi Elahi}, {and} \bibinfo{person}{Christoph Trattner}.} \bibinfo{year}{2023}\natexlab{}.
\newblock \showarticletitle{A survey on popularity bias in recommender systems}.
\newblock \bibinfo{journal}{\emph{arXiv preprint arXiv:2308.01118}} (\bibinfo{year}{2023}).
\newblock


\bibitem[Kowald(2024)]%
        {kowald2024transparency}
\bibfield{author}{\bibinfo{person}{Dominik Kowald}.} \bibinfo{year}{2024}\natexlab{}.
\newblock \showarticletitle{Transparency, Privacy, and Fairness in Recommender Systems}.
\newblock \bibinfo{journal}{\emph{arXiv preprint arXiv:2406.11323}} (\bibinfo{year}{2024}).
\newblock


\bibitem[Kowald et~al\mbox{.}(2013)]%
        {kowald2013social}
\bibfield{author}{\bibinfo{person}{Dominik Kowald}, \bibinfo{person}{Sebastian Dennerlein}, \bibinfo{person}{Dieter Theiler}, \bibinfo{person}{Simon Walk}, {and} \bibinfo{person}{Christoph Trattner}.} \bibinfo{year}{2013}\natexlab{}.
\newblock \showarticletitle{The social semantic server: A framework to provide services on social semantic network data}. In \bibinfo{booktitle}{\emph{9th International Conference on Semantic Systems, I-SEMANTICS 2013}}. CEUR, \bibinfo{pages}{50--54}.
\newblock


\bibitem[Kowald and Lacic(2022)]%
        {kowald2022popularity}
\bibfield{author}{\bibinfo{person}{Dominik Kowald} {and} \bibinfo{person}{Emanuel Lacic}.} \bibinfo{year}{2022}\natexlab{}.
\newblock \showarticletitle{Popularity bias in collaborative filtering-based multimedia recommender systems}. In \bibinfo{booktitle}{\emph{Advances in Bias and Fairness in Information Retrieval, BIAS 2022}}. Springer, \bibinfo{pages}{1--11}.
\newblock
\urldef\tempurl%
\url{https://doi.org/10.1007/978-3-031-09316-6_1}
\showURL{%
\tempurl}


\bibitem[Kowald et~al\mbox{.}(2020a)]%
        {kowald2020utilizing}
\bibfield{author}{\bibinfo{person}{Dominik Kowald}, \bibinfo{person}{Elisabeth Lex}, {and} \bibinfo{person}{Markus Schedl}.} \bibinfo{year}{2020}\natexlab{a}.
\newblock \showarticletitle{Utilizing Human Memory Processes to Model Genre Preferences for Personalized Music Recommendations}. In \bibinfo{booktitle}{\emph{4th Workshop on Transparency and Explainability in Adaptive Systems through User Modeling Grounded in Psychological Theory}}. Association of Computing Machinery.
\newblock
\urldef\tempurl%
\url{https://doi.org/10.48550/arXiv.2003.10699}
\showURL{%
\tempurl}


\bibitem[Kowald et~al\mbox{.}(2023)]%
        {ecir_bias_2023}
\bibfield{author}{\bibinfo{person}{Dominik Kowald}, \bibinfo{person}{Gregor Mayr}, \bibinfo{person}{Markus Schedl}, {and} \bibinfo{person}{Elisabeth Lex}.} \bibinfo{year}{2023}\natexlab{}.
\newblock \showarticletitle{A study on accuracy, miscalibration, and popularity bias in recommendations}. In \bibinfo{booktitle}{\emph{Advances in Bias and Fairness in Information Retrieval, BIAS 2023}}. Springer, \bibinfo{pages}{1--16}.
\newblock


\bibitem[Kowald et~al\mbox{.}(2021)]%
        {kowald2021support}
\bibfield{author}{\bibinfo{person}{Dominik Kowald}, \bibinfo{person}{Peter Muellner}, \bibinfo{person}{Eva Zangerle}, \bibinfo{person}{Christine Bauer}, \bibinfo{person}{Markus Schedl}, {and} \bibinfo{person}{Elisabeth Lex}.} \bibinfo{year}{2021}\natexlab{}.
\newblock \showarticletitle{Support the underground: Characteristics of beyond-mainstream music listeners}.
\newblock \bibinfo{journal}{\emph{EPJ Data Science}} \bibinfo{volume}{10}, \bibinfo{number}{1} (\bibinfo{year}{2021}), \bibinfo{pages}{1--26}.
\newblock


\bibitem[Kowald et~al\mbox{.}(2020b)]%
        {kowald2020unfairness}
\bibfield{author}{\bibinfo{person}{Dominik Kowald}, \bibinfo{person}{Markus Schedl}, {and} \bibinfo{person}{Elisabeth Lex}.} \bibinfo{year}{2020}\natexlab{b}.
\newblock \showarticletitle{The unfairness of popularity bias in music recommendation: A reproducibility study}. In \bibinfo{booktitle}{\emph{Advances in Information Retrieval: 42nd European Conference on IR Research, ECIR 2020}}. Springer, \bibinfo{pages}{35--42}.
\newblock


\bibitem[Lacic et~al\mbox{.}(2022)]%
        {lacic2022drives}
\bibfield{author}{\bibinfo{person}{Emanuel Lacic}, \bibinfo{person}{Leon Fadljevic}, \bibinfo{person}{Franz Weissenboeck}, \bibinfo{person}{Stefanie Lindstaedt}, {and} \bibinfo{person}{Dominik Kowald}.} \bibinfo{year}{2022}\natexlab{}.
\newblock \showarticletitle{What drives readership? an online study on user interface types and popularity bias mitigation in news article recommendations}. In \bibinfo{booktitle}{\emph{European Conference on Information Retrieval}}. Springer, \bibinfo{pages}{172--179}.
\newblock
\urldef\tempurl%
\url{https://doi.org/10.1007/978-3-030-99739-7_20}
\showURL{%
\tempurl}


\bibitem[Lacic et~al\mbox{.}(2015)]%
        {lacic2015utilizing}
\bibfield{author}{\bibinfo{person}{Emanuel Lacic}, \bibinfo{person}{Dominik Kowald}, \bibinfo{person}{Lukas Eberhard}, \bibinfo{person}{Christoph Trattner}, \bibinfo{person}{Denis Parra}, {and} \bibinfo{person}{Leandro~Balby Marinho}.} \bibinfo{year}{2015}\natexlab{}.
\newblock \showarticletitle{Utilizing online social network and location-based data to recommend products and categories in online marketplaces}. In \bibinfo{booktitle}{\emph{Mining, Modeling, and Recommending'Things' in Social Media: Revised Selected Papers of MUSE and MSM'2013 Workshops}}. Springer, \bibinfo{pages}{96--115}.
\newblock


\bibitem[Lex et~al\mbox{.}(2020)]%
        {lex2020modeling}
\bibfield{author}{\bibinfo{person}{Elisabeth Lex}, \bibinfo{person}{Dominik Kowald}, {and} \bibinfo{person}{Markus Schedl}.} \bibinfo{year}{2020}\natexlab{}.
\newblock \showarticletitle{Modeling popularity and temporal drift of music genre preferences}.
\newblock \bibinfo{journal}{\emph{Transactions of the International Society for Music Information Retrieval}} \bibinfo{volume}{3}, \bibinfo{number}{1} (\bibinfo{year}{2020}).
\newblock
\urldef\tempurl%
\url{https://doi.org/10.5334/tismir.39}
\showURL{%
\tempurl}


\bibitem[Lin et~al\mbox{.}(2020)]%
        {lin2020}
\bibfield{author}{\bibinfo{person}{Kun Lin}, \bibinfo{person}{Nasim Sonboli}, \bibinfo{person}{Bamshad Mobasher}, {and} \bibinfo{person}{Robin Burke}.} \bibinfo{year}{2020}\natexlab{}.
\newblock \showarticletitle{Calibration in collaborative filtering recommender systems: A user-centered analysis}. In \bibinfo{booktitle}{\emph{Proceedings of the 31st ACM Conference on Hypertext and Social Media}} \emph{(\bibinfo{series}{HT '20})}. \bibinfo{pages}{197–206}.
\newblock
\showISBNx{9781450370981}


\bibitem[Lops et~al\mbox{.}(2010)]%
        {lops2010content}
\bibfield{author}{\bibinfo{person}{Pasquale Lops}, \bibinfo{person}{Marco De~Gemmis}, {and} \bibinfo{person}{Giovanni Semeraro}.} \bibinfo{year}{2010}\natexlab{}.
\newblock \showarticletitle{Content-based recommender systems: State of the art and trends}.
\newblock \bibinfo{journal}{\emph{Recommender Systems Handbook}} (\bibinfo{year}{2010}), \bibinfo{pages}{73--105}.
\newblock


\bibitem[Luo et~al\mbox{.}(2014)]%
        {luo2014efficient}
\bibfield{author}{\bibinfo{person}{Xin Luo}, \bibinfo{person}{Mengchu Zhou}, \bibinfo{person}{Yunni Xia}, {and} \bibinfo{person}{Qingsheng Zhu}.} \bibinfo{year}{2014}\natexlab{}.
\newblock \showarticletitle{An efficient non-negative matrix-factorization-based approach to collaborative filtering for recommender systems}.
\newblock \bibinfo{journal}{\emph{IEEE Transactions on Industrial Informatics}} \bibinfo{volume}{10}, \bibinfo{number}{2} (\bibinfo{year}{2014}), \bibinfo{pages}{1273--1284}.
\newblock


\bibitem[Mehrabi et~al\mbox{.}(2021)]%
        {mehrabi2021survey}
\bibfield{author}{\bibinfo{person}{Ninareh Mehrabi}, \bibinfo{person}{Fred Morstatter}, \bibinfo{person}{Nripsuta Saxena}, \bibinfo{person}{Kristina Lerman}, {and} \bibinfo{person}{Aram Galstyan}.} \bibinfo{year}{2021}\natexlab{}.
\newblock \showarticletitle{A survey on bias and fairness in machine learning}.
\newblock \bibinfo{journal}{\emph{ACM Computing Surveys (CSUR)}} \bibinfo{volume}{54}, \bibinfo{number}{6} (\bibinfo{year}{2021}), \bibinfo{pages}{1--35}.
\newblock


\bibitem[Moscati et~al\mbox{.}(2023)]%
        {recsys_actr_2023}
\bibfield{author}{\bibinfo{person}{Marta Moscati}, \bibinfo{person}{Christian Wallmann}, \bibinfo{person}{Markus Reiter-Haas}, \bibinfo{person}{Dominik Kowald}, \bibinfo{person}{Elisabeth Lex}, {and} \bibinfo{person}{Markus Schedl}.} \bibinfo{year}{2023}\natexlab{}.
\newblock \showarticletitle{Integrating the ACT-R framework with collaborative filtering for explainable sequential music recommendation}. In \bibinfo{booktitle}{\emph{Proceedings of the 17th ACM Conference on Recommender Systems}}.
\newblock
\urldef\tempurl%
\url{https://doi.org/10.1145/3604915.3608838}
\showURL{%
\tempurl}


\bibitem[M{\"u}llner et~al\mbox{.}(2024)]%
        {mullner2024impact}
\bibfield{author}{\bibinfo{person}{Peter M{\"u}llner}, \bibinfo{person}{Elisabeth Lex}, \bibinfo{person}{Markus Schedl}, {and} \bibinfo{person}{Dominik Kowald}.} \bibinfo{year}{2024}\natexlab{}.
\newblock \showarticletitle{The Impact of Differential Privacy on Recommendation Accuracy and Popularity Bias}. In \bibinfo{booktitle}{\emph{European Conference on Information Retrieval}}. Springer, \bibinfo{pages}{466--482}.
\newblock


\bibitem[Park and Tuzhilin(2008)]%
        {park2008long}
\bibfield{author}{\bibinfo{person}{Yoon-Joo Park} {and} \bibinfo{person}{Alexander Tuzhilin}.} \bibinfo{year}{2008}\natexlab{}.
\newblock \showarticletitle{The long tail of recommender systems and how to leverage it}. In \bibinfo{booktitle}{\emph{Proceedings of the 2008 ACM Conference on Recommender Systems}}. \bibinfo{pages}{11--18}.
\newblock


\bibitem[Resnick et~al\mbox{.}(1994)]%
        {resnick1994grouplens}
\bibfield{author}{\bibinfo{person}{Paul Resnick}, \bibinfo{person}{Neophytos Iacovou}, \bibinfo{person}{Mitesh Suchak}, \bibinfo{person}{Peter Bergstrom}, {and} \bibinfo{person}{John Riedl}.} \bibinfo{year}{1994}\natexlab{}.
\newblock \showarticletitle{Grouplens: An open architecture for collaborative filtering of NetNews}. In \bibinfo{booktitle}{\emph{Proceedings of the 1994 ACM Conference on Computer Supported Cooperative Work}}. \bibinfo{pages}{175--186}.
\newblock


\bibitem[Scher et~al\mbox{.}(2023)]%
        {scher2023modelling}
\bibfield{author}{\bibinfo{person}{Sebastian Scher}, \bibinfo{person}{Simone Kopeinik}, \bibinfo{person}{Andreas Tr{\"u}gler}, {and} \bibinfo{person}{Dominik Kowald}.} \bibinfo{year}{2023}\natexlab{}.
\newblock \showarticletitle{Modelling the long-term fairness dynamics of data-driven targeted help on job seekers}.
\newblock \bibinfo{journal}{\emph{Scientific Reports}} \bibinfo{volume}{13}, \bibinfo{number}{1} (\bibinfo{year}{2023}), \bibinfo{pages}{1727}.
\newblock
\urldef\tempurl%
\url{https://doi.org/10.1038/s41598-023-28874-9}
\showURL{%
\tempurl}


\bibitem[Semmelrock et~al\mbox{.}(2023)]%
        {semmelrock2023reproducibility}
\bibfield{author}{\bibinfo{person}{Harald Semmelrock}, \bibinfo{person}{Simone Kopeinik}, \bibinfo{person}{Dieter Theiler}, \bibinfo{person}{Tony Ross-Hellauer}, {and} \bibinfo{person}{Dominik Kowald}.} \bibinfo{year}{2023}\natexlab{}.
\newblock \showarticletitle{Reproducibility in machine learning-driven research}.
\newblock \bibinfo{journal}{\emph{arXiv preprint arXiv:2307.10320}} (\bibinfo{year}{2023}).
\newblock


\bibitem[Semmelrock et~al\mbox{.}(2024)]%
        {semmelrock2024reproducibility}
\bibfield{author}{\bibinfo{person}{Harald Semmelrock}, \bibinfo{person}{Tony Ross-Hellauer}, \bibinfo{person}{Simone Kopeinik}, \bibinfo{person}{Dieter Theiler}, \bibinfo{person}{Armin Haberl}, \bibinfo{person}{Stefan Thalmann}, {and} \bibinfo{person}{Dominik Kowald}.} \bibinfo{year}{2024}\natexlab{}.
\newblock \showarticletitle{Reproducibility in Machine Learning-based Research: Overview, Barriers and Drivers}.
\newblock \bibinfo{journal}{\emph{arXiv preprint arXiv:2406.14325}} (\bibinfo{year}{2024}).
\newblock


\bibitem[Steck(2018)]%
        {steck2018}
\bibfield{author}{\bibinfo{person}{Harald Steck}.} \bibinfo{year}{2018}\natexlab{}.
\newblock \showarticletitle{Calibrated recommendations}. In \bibinfo{booktitle}{\emph{Proceedings of the 12th ACM Conference on Recommender Systems}} (Vancouver, British Columbia, Canada) \emph{(\bibinfo{series}{RecSys '18})}. \bibinfo{publisher}{Association for Computing Machinery}, \bibinfo{address}{New York, NY, USA}, \bibinfo{pages}{154–162}.
\newblock
\showISBNx{9781450359016}


\end{thebibliography}

\appendix

\end{document}